\documentclass[aps,prl,twocolumn]{revtex4}
\usepackage[dvips]{graphics}
\usepackage[dvips]{graphicx}
\usepackage{amsmath,amsthm,amsfonts,amssymb} 

\begin{document}
\title{Towards an Experimental Test of Gravity-induced Quantum State Reduction.}
\author{Jasper van Wezel$^1$, Tjerk Oosterkamp$^2$ and Jan Zaanen$^1$}
\affiliation{
$^1$Institute-Lorentz for Theoretical Physics, Universiteit Leiden,
P.O. Box 9506, 2300 RA Leiden, The Netherlands \\
$^2$Department of Interface Physics, Kamerlingh Onnes Laboratory, Leiden University, P.O. Box 9504, 2300 RA Leiden, The Netherlands
}
\date{\today}

\begin{abstract}
According to the hypothesis of Penrose and  Di\'osi, quantum state reduction is a manifestation
of the incompatibilty of general relativity and the unitary time evolution of quantum physics~\cite{Penrose96,Diosi89,Ghirardi90}.
Dimensional analysis suggests that Schr\"odinger cat type states should collapse on measurable time scales when masses and lengths of the order of bacterial scales are involved. We analyze
this hypothesis in the context of modern developments in condensed matter- and cold atoms
physics, aimed at realizing macroscopic quantum states. We first consider 'micromechanical'
quantum states, analyzing the capacity of an atomic force microscopy based single spin
detector to measure the gravitational state reduction, but we conclude that it seems impossible to
suppress environmental decoherence to the required degree. We subsequently discuss 'split' cold
atom condensates to find out that these are at present lacking the required mass scale by many
orders of magnitude. We then extent Penrose's analysis to superpositions of mass current carrying
states, and we apply this to the flux quantum bits realized in superconducting circuits. We find
that the flux qubits approach the scale where gravitational state reduction should become
measurable, but bridging the few remaining orders of magnitude appears to be very difficult with
present day technology.
\end{abstract}
\maketitle

\subsection{Introduction} \noindent
The theory of quantum mechanics in principle allows objects of all sizes to appear in spatial superposition states. In nature however, such superpositions are only seen for microscopic particles. There have been a few experiments demonstrating the possibility of creating superposed states of larger objects~\cite{Zeilinger99,Nakamura99,Wal00}, but a truly macroscopic superposition is yet to be observed.
The apparent lack of semi-macroscopic superposition states in nature can be explained to some extent by the influence of environment-induced decoherence~\cite{Zurek81,Zurek82,Joos85}, but ultimately also requires the existence of some mechanism for quantum state reduction~\cite{Adler}.

A seemingly very different realm of physics is in the grip of trying to reconcile Einstein gravity with
quantum theory. These two appear to be mutually exclusive descriptions of reality and according to conventional reasoning
these meet each other at the Planck scale ($10^{-42}$ secs): quantum gravity is supposed to become manifest only very
shortly after the big bang or in the immediate vicinity of a black hole singularity. Not long ago it was suggested by Penrose, and independently by Di\'osi et al., that quantum state reduction might actually be rooted in the troubles between quantum physics and gravity~\cite{Penrose96,Diosi89,Ghirardi90}. If this would indeed be the case it would resolve once and for all the quantum measurement puzzle, while it would also give access to the quantum gravity enigma via manageable experiments.
By dimensional analysis Penrose arrives at the key observation that this gravitational quantum state reduction has to involve mass scales that are beyond the reach of existing technology~\cite{Penrose96}. One typically has to engineer Schr\"odinger cat states from objects having a gravitational mass comparable to that of a large bacterium. Conventional experiments deal with objects that are either very heavy (like humans) or very light (like atoms) compared to this mass scale.

We are of the opinion that the graveness of this subject warrants a concentrated effort to come up with
designs of experiments capable of entering this regime. At present there are two such proposals: Bouwmeester's
mesoscopic mirror experiment~\cite{Marshall03}, and Christian's primordial neutrino oscillations~\cite{Christian05}. Here we will analyze three mainstream
experimental developments aimed at realizing macroscopic quantum superpositions on their potential to be of
use for Penrose's purposes: micromechanical oscillators, cold atom condensates, and superconducting flux
qubits. We do not offer a quick route to success. To the contrary, it turns out that in all these 'paradigms' it
is extremely hard, if not completely impossible, to even get close to the required conditions for observing gravity induced quantum state reduction.

Still, we find it worthwhile to report on these negative outcomes, because we found out that at least in the
examples of the cold atom condensates and the flux qubits, the gravitational state reduction mechanism acts differently in detail from the way originally proposed by Penrose. We hope that this will inspire others to come up with a cleverer experimental scheme than we managed to find.
We start out in section III with an analysis of micromechanical oscillators --systems that form effective illustrations of the gravitational state reduction physics in its simplest setting. In Bouwmeester's experiment with superposed mirrors, it appears that the main bottleneck is to overcome the conventional decoherence coming from the interactions with the environment. As already realized by others, mechanical systems talk too easily to the outside world and this has the consequence that it appears to be very hard to isolate a possible gravitational state reduction mechanism~\cite{Bernad06}. In this regard, both cold atom systems and flux qubits are in a much better shape. They both involve superpositions of superfluid states and these naturally isolate themselves from the environment. The problem with the cold atoms (Section IV) is that at least for experiments that are presently feasible, the factor mass in the quantum state reduction process scales like the square root of the number of particles in the condensate. The number of particles that can be stored in present day condensates is already far too small to cause gravitational state reduction, and the square root scaling makes matters dramatically worse.

In fact, the closest approach to the Penrosian gravitational state reduction regime using proven technology is achieved by the superconducting flux qubits (section V). This also requires most of the real work in this paper. These qubits involve quantum superpositions of current carrying states~\cite{Wal00,Chiorescu00}, and to get a grip on the factor mass in such problems is not straightforward. We believe to have found the right perspective to conceptualize these matters --it is a tricky affair because we lack a full-fledged theory governing the gravity induced quantum state reduction process and one cannot be fully confident regarding the correctness of some underlying assumptions of the heuristic scheme offered by Penrose.  The outcome is a gravitational state reduction time that shows a scaling behavior which differs qualitatively from the elementary case of spatially separated superposed mass distributions. Applying our results to the experimentally realized flux qubits we find that these come tantalizingly close to the Penrosian regime. A flux qubit optimized to probe the gravitational state reduction should be capable of undergoing gravitational collapse on the second time scale. Unfortunately, there seems to be a fundamental border coming from the environmental decoherence at time scales of milliseconds. It appears to be impossible for fundamental reasons to bridge these remaining 3-4 orders of magnitude in time required for the flux qubits to detect gravitational state reduction. But before we analyze these various examples in detail, let us first present some generalities regarding Penrose's state reduction ideas.

\section{General background.}  \noindent
The idea of gravitational quantum state reduction~\cite{Penrose96,Diosi89,Ghirardi90}  is  based on the incompatibility of general covariance and unitarity~\cite{Penrose96}. In the absence of a theory of quantum gravity, it is just asserted that unitarity comes
to an end when the effects of covariance become noticeable and the surprise is that this hypothesis leads to measurable
consequences under remarkably mundane conditions.  It can be argued that a characteristic energy scale is associated with
the impossibility to assign a pointwise identification of space times characterized by a mass at two different spatial positions.
This is the gravitational self-energy, in the non-relativistic (Newtonian) case\cite{Penrose96,Christian05}:
\begin{eqnarray}
\Delta E = \xi G \int \int \frac{ \left[ \rho \left( {\bf x} \right) - \rho' \left( {\bf x} \right) \right] \left[ \rho \left( {\bf y} \right) - \rho' \left( {\bf y} \right) \right] }{\left| {\bf x} - {\bf y} \right|} d{\bf x} d{\bf y}.
\label{DeltaE}
\end{eqnarray}
Here $G$ is Newton's gravitational constant while $\rho$ is the mass density of one member of the superposition, $\rho'$ is the mass density the other member, and $\xi$ is a dimensionless parameter that can be interpreted as the "strength" of the quantum state reduction process~\cite{Christian05}, which is expected to be of order unity. Balancing this gravitational self energy with
Planck's constant gives rise to a 'Planckian' time $\tau$,
\begin{eqnarray}
\tau = \frac {\hbar}{\Delta E}.
\label{tau}
\end{eqnarray}
that is associated with the characteristic time it takes for the quantum superposition to be reduced to only one of its member states. In the case of a mass $M$, in coherent superposition with itself displaced over a distance that is very large compared to the spatial extent $L$ of its mass distribution, one can get away with just counting the dimensions of Eq.'s~\eqref{DeltaE} and~\eqref{tau},
\begin{eqnarray}
\tau_0 = \frac{\hslash L}{\xi G M^2}
\label{dimensionaltau}
\end{eqnarray}
For instance, consider the electron mass and $L=10^{-10}$m, the characteristic dimensions of an electron in an atom, and using
$\xi = 1$ it follows
that $\tau \simeq 10^{26}$ secs., i.e. atoms are stable against gravitational state reduction for a very long time indeed~\cite{Penrose96}. On the other hand, the typical feline dimensions of  $M =1$ kg and $L=1$ m imply that the famous Schr\"odinger cat can
only be simultaneously dead and alive during  approximately $10^{-24}$ secs. The known empirical facts  for macroscopic and microscopic
objects are clearly not inconsistent with Penrose's assertion. To critically test the hypothesis one would therefore need to
design experiments that take the system in a controlled way through the gravitational state reduction regime. Consider a
feasible time of a second, and a mass of $10^{-15}$ grams (mass of e-coli): the required length scale that follows from~\eqref{dimensionaltau} is $L \simeq 0.5~\mu$m. The rules of physics
do not at all forbid such an experiment, but it would obviously require a breakthrough in 'nano-engineering'.

Although a main aim of this paper is to demonstrate circumstances where it fails, this simple dimensional consideration gives away why Penrose's assertion still goes untested. Another aim of this paper is to make clear how far away several existing experimental schools of
thought are removed from realizing an experimental test of Penrose's idea. There is no way around the basic requirement that one has to form superpositions of
considerable masses over considerable distances. It is very hard to shield such systems from environmental states that cause
conventional decoherence. According to Penrose's hypothesis the gravitational state reduction in a state that is highly entangled with its environment would be no less active but it would be impossible to retrieve unambiguous information regarding the workings of the quantum state reduction from an effective description of such a messy superposition.

There have been a few previous proposals of experimental tests for the idea of gravity-induced quantum state reduction~\cite{Christian05,Penrose98,Marshall03}. One example of an experiment that uses well-shielded particles is the proposal by Christian~\cite{Christian05} to look for shifts in the flavor ratios of cosmogenic neutrinos. Because neutrinos interact only via gravitational and weak forces, they can remain quantum coherent while traveling over cosmological distances. According to equation~\eqref{tau} these microscopic particles should have a large but finite lifetime just like any other superposition state that involves gravitationally distinct parts of its wavefunction. In the case of cosmogenic neutrinos, Christian showed that the gravitation-induced quantum state reduction will lead to a shift in the flavor oscillations of the neutrino, and thus to an observable departure from the expected flavor ratios of neutrinos observed on earth~\cite{Christian05}. The actual observation of these flavor ratios however, awaits the construction of a number of neutrino detectors~\cite{Christian05,Beacom03}.

An example of the other route toward a measurable quantum state reduction --producing a very massive superposition state and trying to shield it from external sources of decoherence-- is the proposal by Bouwmeester et al. to use a small mirror with a mass of order $10^{-12}$ kg mounted on a very weak spring at the end of an optical cavity~\cite{Marshall03}. By placing the optical cavity in one arm of a Michelson interferometer, the mirror can be brought into a state of spatial superposition: if the photon is in the one arm, then the mirror will be displaced by the optical pressure, whereas it will remain in its original position if the photon is in the other arm. Due to the requirements on stability, quality factor and temperature of the setup however~\cite{Marshall03}, the experiment has not actually been performed yet, and doubts have been expressed as related to the fundamental difficulties associated with isolating mechanical
objects like mirrors from the environment~\cite{Bernad06}.

Finally, a caveat: since the gravitational state reduction of the 'normal' (not neutrino) type  requires that microscopic entities displace collectively as one whole macroscopic object,  one is necessarily dealing with rigid matter characterized by an order parameter. It is a subtle fact
that such matter for the very reason that it is rigid is characterized by an intrinsic decoherence time $\tau = N \hbar / (k_B T)$
where $N$ is the number of constituent particles\cite{vanWezel05,vanWezel06}. This is surely setting a lower bound to the size of the gravitating body.

\subsection{The Micromechanical Oscillator} \noindent
In recent years there has been great experimental progress in constructing and monitoring microscopic mechanical oscillators~\cite{Zalalutdinov00,Knobel03,Sazonova04}. One example is the experiment by Sazonova et al. in which a carbon nanotube is suspended from its ends and in which the vibrational modes of the so formed string can be measured using the electronic properties of the carbon nanotube itself~\cite{Sazonova04}. By attaching a microscopic piece of magnetic material to the string one could create a fluctuating magnetic field associated with the string's vibrational modes. This magnetic field can then be read out using a SQUID as a pick-up loop. Conversely one can also use the coupling between the vibrations and the current states of the SQUID to impose a quantum superposition of vibrational modes using a superposition of different flux states in the SQUID. Because of the relatively low mass and short length of the vibrating carbon nanotube, the superpositions of its vibrational modes will take a rather long time to collapse, and therefore it seems unlikely that this type of experiment could be directly used to measure the timescale proposed by Penrose. It may be possible however, to use variations of this principle of inducing superposition states onto microscopic mechanical oscillators to reach the desired timescale.


In general one needs to couple a mechanical oscillator to a suitable qubit state which can exert a force onto it, in order to induce a quantum superposition state in the resonator. The first example of an experiment which aims to measure gravity-induced collapse using a mechanical oscillator is the one proposed by Bouwmeester et al.~\cite{Marshall03}. In that case the qubit state is provided by a photon in a high finesse cavity which bounces off a mirror that is mounted on a small cantilever. The photon will bounce around inside the cavity for a time of order $Q L /c$, with $Q$ the quality factor of the cavity, $L$ its length and $c$ the speed of light. In Bouwmeester's experiment $Q=10^5$ and $L=1$~cm, so that the photon will be exerting a force onto the resonator for about $3$~$\mu$s. For a cantilever with a mechanical resonance frequency of say $\omega=100$~kHz this time would be long enough for the cantilever to respond to the presence of the photon. The expected displacement of the mirror on the cantilever can then be estimated as $d=F/k$ with $F$ the force exerted by the photon, and $k$ the spring constant of the cantilever. Using $F=(\hbar c )/ (\lambda L )$ and $k=M \omega^2$, one can then easily see that a cubic mirror with $10$~$\mu$m edges will be displaced by the optical pressure over a distance which is about equal to its quantum mechanical zero-point motion $\sqrt{\hbar/(M \omega)}$. Using this distance, Bouwmeester estimated that with the current state of the art technology the gravity-induced reduction time would be of the order of $10^3$~sec, which is still six orders of magnitude larger than the environmental decoherence time~\cite{Marshall03}.

\begin{figure}[hbt]
\center
\includegraphics[width=0.8\columnwidth]{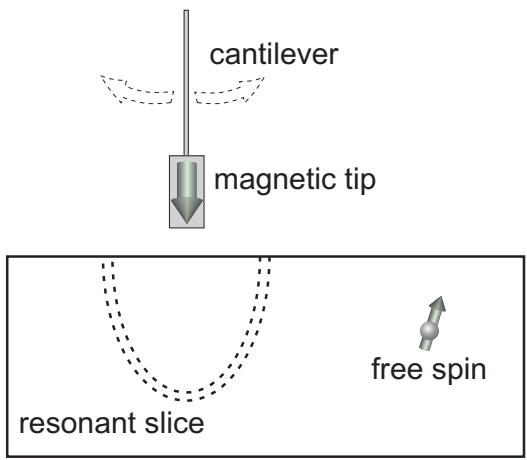}
\caption{A schematic overview of the AFM setup used by Rugar et al.~\cite{Rugar04} The combined magnetic field from the cantilever tip and the external field creates a spin-resonance slice which is swept across a silicon surface with a low density of unpaired spins.}
\label{spinDetection}
\end{figure}

A second realizable experimental setup using a microscopic mechanical oscillator which could in principle be used to measure the gravity-induced quantum state reduction, is the spin detection experiment of Rugar et al.~\cite{Rugar04, Mamin03}. In this setup a magnetic tip is attached to a small AFM cantilever. By combining the magnetic field of the tip with a static external magnetic field, a thin region is created in the material under the cantilever in which the total field satisfies the condition for electron spin resonance (see figure~\ref{spinDetection}). When the tip is swept across an area of silicon surface with a single unpaired spin, then the spin will feel a resonant magnetic field each time the resonant slice passes through its position. The resonant field will cause the spin to flip while at the same time the spin exerts an alternating magnetic force on the cantilever~\cite{Rugar04}. The force exerted on the cantilever by the spin causes a shift in the vibration frequency of the cantilever and consequently allows the detection of the spin by the cantilever.

In their experiment, Rugar et al. assume that the cantilever is enough of a 'measurement machine' to instantaneously collapse the unpaired spin into an eigenstate which is either aligned or anti-aligned with the motion of the cantilever~\cite{Rugar04}. In principle though, it should also be possible to prepare the unpaired spin in a superposition state at a moments when the cantilever is far away from the position in which the resonant field condition is satisfied. According to Penrose's proposal then, the spin will only collapse when its coupling to the cantilever motion has caused enough mass to be included into the superposition. Because the different alignments of the spin states have an opposite effect on the motion of the cantilever, the spin superposition would cause the cantilever to adopt a superposition of different vibrational states. The effect of the spin on the cantilever can be easily estimated by looking at the force between the cantilever and the spin. This force is just $F=\mu_B \partial B/ \partial x$, with $B$ the lateral magnetic field and $\mu_B$ the magnetic moment of the electron. Substituting the values used in the experimental setup of Rugar et al. we find that the force exerted is about $2$~aN. If the cantilever were at rest in its equilibrium position, then this force corresponds to a displacement of $d_0=F/k \simeq 2$~fm (here $k$ is again the spring constant of the cantilever).

\begin{figure}[hbt]
\center
\includegraphics[width=0.8\columnwidth]{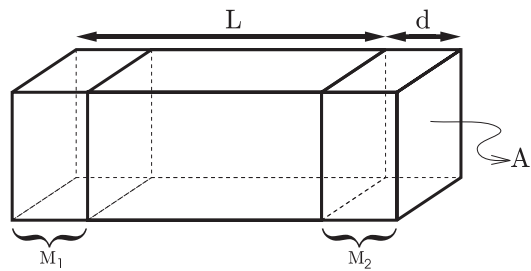}
\caption{Schematic depiction of the superposition state of the oscillator. Both copies of the oscillator have length $L$, and have moved $d/2$ from their coincidence position.}
\label{current}
\end{figure}

We can gain a lot of extra displacement by using the $Q$ factor of the cantilever. If the cantilever is allowed to vibrate back and forth over the superposed spin, then the amplitude difference between the superposed vibrational states which is built up over $Q$ oscillations is $d=d_0 Q \simeq 1$~\AA. Obviously this displacement is much smaller than the actual thickness of the cantilever tip. The two superposed copies of the oscillator will thus always partially overlap, and to estimate the gravity-induced collapse time in such a situation we cannot straightforwardly apply equation~\eqref{dimensionaltau}.
Instead we start by approximating the shape of the oscillator to be that of a bar with length $L$ and surface area $A$ (see figure~\ref{current}). The overlapping piece of the two copies of the oscillator does not contribute to the self energy of equation~\eqref{DeltaE}, and the self energy is completely determined by the interaction between the non-overlapping flanks of the two superposed bars. The distance between the centres of mass of these flanks is always just $L$, so that the gravitational interaction between them can be estimated as
\begin{eqnarray}
\Delta E &=& \xi \frac{1}{2} G \frac{M_1 M_2}{L} \nonumber \\
&=& \xi \frac{1}{2} G M^2 \frac{d^2}{L^3},
\label{DeltaE1}
\end{eqnarray}
in which $M$ is the mass of the total micromechanical oscillator, $M_{1}$ and $M_2$ are the masses of the flanks, and $d$ is the distance that the oscillating copy has moved from its rest position. The factor $1/2$ comes from the relative weight that each copy has in the superposition's wavefunction.

Ideally, to be able to observe gravity-induced quantum state reduction in the spin detection setup, the timescale $\tau$ implied by eq.~\eqref{DeltaE1} should be smaller than the period of oscillation which is set by the oscillator's eigenfrequency $\omega$.
Even if we approximate $\Delta E$ to be constant during an entire period and equal to its maximum value (which corresponds to assuming $d=1$~\AA~throughout), then still this implies for a silicon oscillator of dimensions $10$~$\mu$m~x~$10$~$\mu$m~x~$100$~nm that its gravitational collapse time is of the order of $10^4$ sec.

This time is clearly much too long to be measured in the current spin detection setups. The experiment can be optimized beyond the current setup using a stronger magnetic field gradient, a weaker spring constant and different oscillator dimensions. This could in principle lead to a decrease of the expected gravity-induced quantum state reduction time of several orders of magnitude. The greatest problem in actually registering the gravity-induced effect however, is not the mere size of the collapse time, but rather the distinction between this effect and the environment-induced decoherence, or even the decoherence induced by the photons measuring the displacement of the cantilever. Because the cantilever is a mesoscopic crystal, it is susceptible to all the usual sources of decoherence. Disentangling the gravitational collapse from the environmental decoherence mechanisms will be the true challenge in experiments based on micromechanical oscillators.

\section{The cold atom Bose-Einstein Condensate} \noindent
The Bose-Einstein condensate formed in cold atom gasses is a semi-macroscopic, coherent state of matter that can nowadays be routinely produced and manipulated~\cite{Anderson95,Bradley95,Davis95}. Using spectroscopy and interferometry experiments, the collective condensate wavefunction (consisting of order $10^5$ atoms), can be brought into a spatial superposition state. These experiments can be done while the condensate is freely falling, while it is in a magneto-optical trap, and even on the surface of a chip~\cite{Giltner95,Stenger99,Kozuma99:PRL,Kozuma99,Torii00,Xiong02,Muller03,Wang05}. Since these condensates are formed from
rather heavy objects (the atoms) while they are effectively very well screened from the environment, the question arises if these can be used for Penrosian purposes. Since the quantum coherence can be easily tracked by interferometry experiments, gravitational
quantum state reduction would be easily detectable. By varying the number and the mass of the atoms one could easily distinguish the
gravitational effects from other sources of decoherence. In other words, this appears at first sight to be an ideal stage for Penrosian
experiments; but as we will now discuss the factor mass is not quite cooperative.

By combining several cold atom wizardry's that have been recently reported~\cite{Xiong02,Torii00,Wang05} the following
experiment comes to mind.  Using  Bragg diffraction of a Bose Einstein condensate against a moving, optical standing wave
one can create a quantum coherent state consisting of two parts moving in opposite directions in space.
This Bragg diffraction can be understood as a two-photon stimulated Raman process, where the atomic momentum is changed by coherently scattering a photon from one laser beam into another~\cite{Giltner95,Stenger99,Kozuma99:PRL}. It can be used to couple the zero momentum condensate state $\left| p=0 \right>$ to the 1$^{\text st}$ order Bragg diffracted states $\left| p = \pm 2 \hbar k \right>$ (with $k=2\pi / \lambda$ and $\lambda$ the laser wavelength)~\cite{Stenger99,Kozuma99:PRL,Kozuma99}. The state of the Bose-Einstein condensate after the diffraction has occurred is of the schematic form
\begin{eqnarray}
\left| \Psi \right> = \left( \psi^{\dagger}_{L} + e^{i \theta} \psi^{\dagger}_{R} \right)^N \left| \text{vac} \right>,
\label{BECstate}
\end{eqnarray}
where $\psi^{\dagger}_{L}$ ($\psi^{\dagger}_{R}$) creates an atom in the left (right) moving cloud, $N$ is the total number of atoms in the condensate, and the relative phase $\theta$ is set by the phase of the standing wave in the middle of the initial wavepacket~\cite{Leggett91,Torii00}. The effects  of the interactions are ignored but these can be accounted for perturbatively and are inconsequential in the present context.

This state is not quite the superposition state considered by Penrose and Di\'osi, which had the form $[(\psi^{\dagger}_{L})^N + (\psi^{\dagger}_{R} )^N] \left| \text{vac} \right>$. For gravitational purposes, the difference is that not $N$ particles are present simultaneously at two distinct position in space, but instead one has to focus in on the typical mass fluctuation associated with the phase coherent state Eq.~\eqref{BECstate}. It is well understood~\cite{Sols98,Javanainen99} that in this coherent state a fraction $\sqrt{N}$ of the atom masses are simultaneously present in the spatially separated subcondensates $L,R$:  the $\sqrt{N}$ appears here because it is the standard deviation of the difference in particle number between the two sides in the binomial distribution of equation~\eqref{BECstate}). Since all that matters for the gravitational state reduction is that this fraction $\sqrt{N}$ of the atoms are simultaneously at two different locations in
space, the state Eq.~\eqref{BECstate} should still fall prey to gravitation after a time $\tau=\hbar / \Delta E$ with $\Delta E$ determined by the mass fluctuations. After the gravitational state reduction the difference in particle number between the two parts of the condensate should acquire a definite value, because these number states are the only ones that are gravitationally stable. The reduction from a coherent superposition state to a definite number state can be seen experimentally as the disappearance of phase-coherence between the different parts of the condensate. The phase-coherence can be measured by bringing the two parts of the condensate back together, either by putting them in a very shallow magnetic trap as in ref.~\cite{Xiong02} or by applying a $\pi$-pulse with the Bragg beams as in refs.~\cite{Torii00,Wang05}. Once the two parts of the condensate are back together, one can apply another Bragg pulse and study their interference pattern as a function of the applied phase shift $\theta$ or an applied magnetic field gradient~\cite{Torii00,Wang05}.

To estimate the gravitational self energy $\Delta E$ we will approximate the shape of the Bose Einstein condensate to be that of a bar with length $L$ and surface area $A$ (see figure~\ref{current}). At time $t=0$ the two copies of the condensate bar will start moving apart, but like in the case of the micromechanical oscillator, each copy will still partially overlap with its partner until $t=t_1$, at which point the outer ends of the two copies just touch. As before, the self energy prior to $t_1$ is completely determined by the interaction between the non-overlapping flanks of the two condensate bars (see figure~\ref{current}), so that the gravitational interaction between the two quantum copies is of order
\begin{eqnarray}
\Delta E \left( t \leq t_1\right) &=& \xi \frac{1}{2} G \frac{M_1 M_2}{L} \nonumber \\
&=& \xi \frac{1}{2} G M^2 \frac{\left(2vt\right)^2}{L^3},
\label{DeltaE2}
\end{eqnarray}
where $M$ is the mass of the total condensate, $M_{1}$ and $M_2$ are the masses of the flanks, and $v$ is the speed at which the condensates move apart. After the time $t=t_1$, the two copies of the condensate are completely detached from each other, and then their Newtonian gravitational self energy is approximately given by
\begin{eqnarray}
\Delta E \left( t \geq t_1\right) &=& \Delta E \left(t_1 \right) + \xi \frac{1}{2} G M^2 \left[\frac{1}{L}-\frac{1}{2vt}\right] \nonumber \\
&=& \xi G \left(\rho A\right)^2 \left[L-\frac{L^2}{4vt}\right],
\label{DeltaE3}
\end{eqnarray}
where in the last line we used $M=\rho A L$, with $\rho$ the mass density of the condensate, so that everything can be expressed in terms of the length of the condensate bar.

As in the case of the micromechanical oscillator, the timescale $\tau$ as determined by the gravitational self energy $\Delta E$ must be smaller then the time it takes for the two copies of the condensate to recombine in order for the gravity-induced quantum state reduction to become manifest. In addition, the time $\tau$ should also be smaller than the decoherence time of the Bose-Einstein condensate due to environmental influences. Coherence times of up to the order of $100$ ms can be achieved in cold atomic gasses using state of the art techniques, while in the experiments with Bragg scattered Bose Einstein condensates oscillation periods of about $10$ ms have been reported~\cite{Torii00,Xiong02}.

Let us consider a best case scenario, corresponding with the oscillation period being equal to the maximal coherence time of $100$ ms. Let us even approximate the self energy of the separating copies of the condensate to correspond with its maximum possible value,
ignoring the effects of the overlap of the quantum copies. The simple estimate Eq.~\eqref{dimensionaltau} for the gravitational state reduction time can then be used with $M = \sqrt{N} m_{at}$:
\begin{eqnarray}
\tau_{\text{min}} = \hslash / \Delta E_{\text{max}} = \frac{\hslash L}{\xi G M^2} \nonumber \\
\tau_{\text{min}} < 100~ms \Rightarrow \xi M^2 > 10 \frac{\hslash L}{G}.
\end{eqnarray}
to give some idea about the numbers let us consider a condensate formed from Rubidium atoms with a density of order $10^{-2}$ kg/m$^3$, and an isotropic spatial extent $L=\sqrt[3]{N m_{at} / \rho}$. Let us again assume that $\xi = 1$ and it follows that the condensate has to be formed from a total of order $10^{29}$ atoms, a number
exceeding by a factor of $10^{23}$ the maximum condensate size that is at present technologically feasible -- of course it is anyway not very likely that the
required huge condensates could remain isolated from the environment during the time of the experiment.

We thus find that in this set-up using proven technology the atomic condensates perform much worse than the simple mechanical oscillator as a
test bed for gravitational state reduction. The reason resides clearly in the unfavorable scaling associated with the fact that only
$\sqrt{N}$ of all atoms are involved in the gravitational effects. In fact, the prospects would be much brighter if it would appear
possible to generate condensate Schr\"odinger cat states of the form  $[(\psi^{\dagger}_{L})^N + (\psi^{\dagger}_{R} )^N] \left| \text{vac} \right>$. The mass scaling would then involve $N$ instead of $\sqrt{N}$. Using the simple estimate Eq.~\eqref{dimensionaltau}
and taking the other conditions the same as in the above one would find that only $10^{11}$ Rubidium atoms are needed. To the best of
our knowledge, however, it is at present unknown how to generate such superpositions. Starting with a single condensate the only
way one can impose the entanglement on number states of atoms is by borrowing the quantum information from the radiation field.
Unfortunately, the amplitude of radiation fields formed from entangled photons is too small by many orders of magnitude for it to be used to form quantum copies of entire cold atom clouds drifting apart in space as a whole.

\section{Flux quantum bits and gravitational state reduction} \noindent
This section deals with a subject that is conceptually quite different from the straightforward examples discussed in the previous sections, and it is likely also the most controversial one. The question we address is: are coherent superpositions formed from semi-macroscopic
states that differ in {\em mass current} instead of {\em mass density} influenced by gravitational state reduction? Empirically
this is an important issue: states of superposed currents are now routinely produced in superconducting flux qubits ~\cite{Wal00,Chiorescu00} where currents of the order of a micro amp\`ere (involving up to $10^{11}$ electrons) are brought in coherent superposition of running both clockwise and counter clockwise. These superposition states can stay quantum coherent for rather long times like $10^{-7}$ seconds, and they can be viewed as benchmark Schr\"odinger cat states.

A priori it is not at all obvious whether these currents are affected by gravitational effects. Consider the static mass distributions of the electrons in the two fluxoid states with opposite chirality: these are actually precisely identical. Applying the Penrose recipe naively one would thus infer that there are no gravitational self energies at work at all because the mass distributions associated with the two current states are indistinguishable. One could then worry about gravitational framedragging effects, but these contain an extra prefactor $v/c$ and since the superfluid velocity $v$ is tiny compared to the light velocity $c$ one will arrive at the conclusion that the flux qubits are for every practical purpose shielded from the unitarity-covariance conflict.

Driven to the extreme however, this consideration leads to paradoxical conclusions, as can be illustrated by a simple gedanken
experiment. Imagine a macroscopic metal ring with a weight in the kilogram range that is engineered to be perfectly round with nanometer specifications, so that the change in mass distribution caused by a rotation of the  ring  is very small as compared to the characteristic
gravitational reduction mass. This ring is suspended in, say, intergalactic space and cooled to nanokelvins to shield it from environmental influences. The above reasoning would imply the following nonsensical outcome: when by some advanced quantum engineering this
ring would be prepared in a coherent superposition of some initial state and state in which it has undergone a pure counterclockwise rotation of f.e. 90 degrees, then this state would live forever.
Why is this nonsensical? Divide the ring  into small segments, each weighing say a gram. In the rotated ring each of these segment will have moved to
the left relative to its position in the static ring. Because these segments each way a gram this implies that the superposition state of the ring as a whole involves clearly separated superposition states of objects of everyday mass scales. Clearly such superposition states should be the first to fall prey to gravitational state reduction if Penrose's hypothesis is fit. The coherence of the parts is
a necessary condition for the coherence of the whole and because the segments cannot sustain a quantum superposition, the state of the ring as a whole will also be reduced.

In the original formulation by Penrose~\cite{Penrose96, Penrose98} the need for a condition of coherence of all parts is not at all obvious, but implicitly it has always been clear that one must only consider small identifiable parts, rather than look only at the whole. After all, a superposition of a single electron over a distance of $10^{-10}$ m is not supposed to be reduced within any measurable time. But if one simultaneously considers all the electrons in the universe (which after all make up only a single spacetime metric for the universe as a whole), then surely together they will have a gravitational self energy corresponding to a truly minute decay time $\tau$. Certainly this is not the way quantum state reduction works. All of the electron superpositions can be treated as truly separate systems, with individual reduction times that can be calculated as if each superposition makes up its own individual spacetime metric.

The only way in which one superposed electron state could influence another is if the two together formed an {\em entangled} state. In that case standard lore teaches us that the reduction of the state of one part of the wavefunction will automatically also reduce the other part. Even in this case though, the idea that an entangled state of two parts would be more robust against quantum state reduction then each of its constituents individually seems very hard to believe. In the absence of a full fledged theory of quantum state reduction giving a clear explanation of such a marvelous reduction-cancellation construction, we will just assume that it does not exist and stick to the rule that if the parts of a superposition do not want to be coherent, one can forget about the coherence of the whole.

Although it might not be immediately obvious, the above ring in outer space serves as an ideal metaphor for countercirculating supercurrents. To appreciate this, one has to recollect some well understood facts about the hydrodynamics of superfluids. Standard (classical) hydrodynamics starts out with the notion of the 'element of fluid', i.e. one can subdivide the flow in small volumes, ascribing a fixed number of microscopic constituents to every fluid element. The equations of  hydrodynamics then tell us how the collective flow constituted from these indivisible units evolves in time; e.g. one can make the history of these fluid elements visible by adding tracers to the flow and follow the streamlines. These elements of fluid are as good as the segments of the ring discussed in the above for the purpose of identifying how the microscopic constituents are transported; this hangs together with conservation of number and compressional rigidity, shared by
the fluid and the solid, and is unrelated to the shear rigidity special to the solid.

It is a well established physical principle that macroscopic superflow as found in the flux qubits is also of the
hydrodynamical nature; it is just like classical hydrodynamics except that superflow is irrotational and thereby dissipationless. Henceforth, one can also divide the superflow into elements of fluid and these elements of fluid take the role of the segments of the ring. Surely, one has to take care of the microscopic cut-offs that are different from the normal fluid: one cannot meaningfully define an element of fluid  at distances shorter than either the coherence length (size of the Cooper pair) or the length associated with the number-phase uncertainty relation  ( $\Delta \varphi \Delta N > \hslash$ and $\Delta \varphi \propto 1/\sqrt{N}$).

Based on these observations we will now develop a recipe to determine when gravitational collapse occurs in coherent superpositions of current carrying states. This will work a bit differently from the straightforward mass distributions discussed before, with the effect that the elementary relation
Eq.~\eqref{dimensionaltau} is actually seriously violated.

\subsection{linear current distributions}  \noindent
Before we turn to the description of an experiment using real superconducting flux qubits, let us first examine the easier (and purely theoretical) case of a superposition of two counterpropagating linear currents in an infinitely long, straight wire.
At time $t=0$ we can identify a piece of supercurrent of length $L$, and then follow it along as it starts to flow both to the left and to the right. In the beginning the two superposed copies of the original piece will still overlap, and the gravitational self energy of their difference in mass distribution will be due only to the non-overlapping flanks. At a time $t=t_1=L/(2v)$, the two copies will just touch each other, and then start to move apart, thus increasing their gravitational energy because of a growing distance, rather than because of a growing amount of non-overlapping mass. As in the case of the Bose-Einstein condensate, the gravitational self energy of the piece of supercurrent is easily found to be:
\begin{eqnarray}
\Delta E \left( t \leq L/(2v)\right) &=& \xi \frac{1}{2} G \left(\rho A\right)^2 \frac{\left(2vt\right)^2}{L} \nonumber \\
\Delta E \left( t \geq L/(2v)\right) &=& \xi G \left(\rho A\right)^2 \left[L-\frac{L^2}{4vt}\right],
\label{DeltaE4}
\end{eqnarray}
where $\rho$ is the mass density of the supercurrent, $v$ its speed and $A$ the cross section of the wire orthogonal to the direction of flow.
Using equation~\eqref{tau} the self energy $\Delta E$ can be converted into the timescale $\tau$ at which the effects of gravity are expected to  become noticeable. The actual quantum state reduction of the state of superposed pieces of current will then occur as soon as the timescale $\tau$ equals the actual time that the superposition has existed. Using equations~\eqref{DeltaE4} this yields the timescale $t_{\text{QSR}}$ for the quantum state reduction of a patch of supercurrent of length $L$:
\begin{eqnarray}
t_{\text{QSR}} &=& \tau\left( t_{\text{QSR}} \right) = \frac{\hslash}{\Delta E \left( t_{\text{QSR}} \right)} \nonumber \\
\Rightarrow t_{\text{QSR}} &=& \left\{ \begin{array}{cc} \sqrt[3]{\frac{\hslash L}{2 \xi G \left(\rho A v\right)^2}} & \text{if } ~t_{\text{QSR}} \leq \frac{L}{2v} \\ & \\ \frac{ \hslash}{\xi G \left( \rho A \right)^2 L} +\frac{L}{4 v} & \text{if } ~t_{\text{QSR}} \geq \frac{L}{2v} \end{array} \right. .
\label{QSR}
\end{eqnarray}

\begin{figure}[hbt]
\center
\includegraphics[width=0.6\columnwidth]{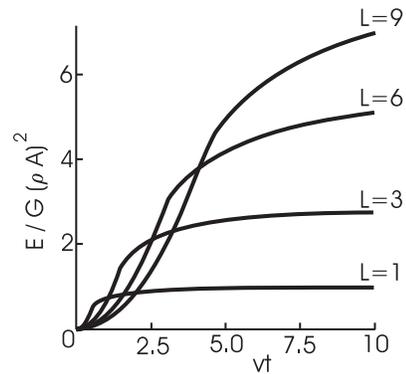}
\caption{Sketch of the gravitational self energy of the difference in mass distribution of two separating copies of a block of supercurrent in a linear chain. The different curves represent different sizes of the block of supercurrent under consideration.}
\label{LinCurrent}
\end{figure}

The entire superposition state of counterpropagating currents in the infinitely long wire can remain coherent only as long as all of its individually identifiable parts can remain in a coherent superposition state by themselves. To find the time at which the total state of the entire system will be reduced, we thus consider all possible sizes $L$ of our piece of supercurrent, and identify the one with the lowest corresponding $t_{\text{QSR}}$ as being responsible for the entire quantum state reduction. The minimum reduction time as a function of $L$, and the corresponding size of the piece of supercurrent responsible for the quantum state reduction are easily seen from~\eqref{QSR} to be
\begin{eqnarray}
t_{\text{QSR}}^{\text{min}} &=& \sqrt{\frac{\hslash}{\xi G \left( \rho A \right)^2 v} } \nonumber \\
L_{\text{QSR}} &=& \sqrt{\frac{4 \hslash v}{\xi G \left( \rho A \right)^2 v} }.
\label{mins}
\end{eqnarray}
Notice that $L_{\text{QSR}} = 2 v t_{\text{QSR}}^{\text{min}}$, which means that the quantum state reduction happens exactly at the moment of separation of the two copies of length $L_{\text{QSR}}$.
The difference in scaling between $\tau$ in~\eqref{dimensionaltau} and this timescale~\eqref{mins} is due to the fact that the current state consists of a combination of all possible sizes of constituent segments. Its collapse time is automatically minimized with respect to the segment size, whereas the elementary superposition of a single mass distribution over two distinct positions has a well defined (maximum) size of its own.

Because of the requirement that the whole cannot be coherent unless all parts can be coherent, we thus find that a superposition of counterpropagating supercurrents in an infinitely long wire will undergo gravity-induced quantum state reduction at the time $t_{\text{QSR}}^{\text{min}}$. This time is not infinitely long, as could be expected when blindly applying Penrose's formalism, nor is it infinitely short as could be expected from the realization that after any infinitesimal time an infinite amount of mass will be in a quantum superposition state.

If one could actually construct an infinitely long superconducting wire with a cross section comparable to those used in the flux qubits, and one could somehow create a current of, say, about $1$ $\mu A$ running both up and down the wire, then this superposed current could exist according to~\eqref{mins} for just about one hour. In comparison, a human moving into a superposition state with a speed of $1$ m/s would collapse after $10^{-14}$ seconds, and reach a maximum separation of superposed copies of only $10^{-14}$ meters.

The described mechanism for finding the time at which a superposition of counterflowing current states will be reduced is generic for all such current states. The geometry of the currents may add some prefactors and the topology of the setup may give rise to additional constraints, but ultimately these are all just variations on the theme. The deciding factor in our analysis is the assertion that the current as a whole can only stay quantum coherent for as long as each and every one of its constituent parts can sustain a coherent superposition state.

In comparison with the elementary setup of superposing only a single mass, the use of current states does not offer a more approachable reduction time: the timescale~\eqref{mins} may be small, but it is of course never smaller than the time that could be obtained by making a direct superposition state using only the segment $L_{\text{QSR}}$. On the other hand, using superconducting supercurrents rather than isolated particles does have the profound advantage that these states can be much more easily screened from decoherence. Going through the trouble of using these more complicated objects may therefore still be a price worth paying.

\subsection{The Superconducting Flux Qubit} 
\begin{figure}[hbt]
\center
\includegraphics[width=0.3\columnwidth]{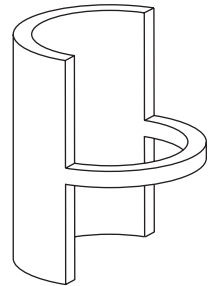}
\caption{Schematic picture of the proposed flux qubit geometry. By independently varying the cross sectional areas of the two regions it is possible to study the dependence of the collapse time on the geometry of the qubit.}
\label{tube}
\end{figure}

\noindent
Having worked out the simple case for a straight, infinitely long superconducting wire, we can now turn to the actual flux qubit. Eventually, the main experimental difficulty will be to distinguish the gravitationally induced quantum state reduction from the standard environmental decoherence processes~\cite{CaldeiraLeggett}. One possible way to do so is to see if the reduction time depends on the exact geometry of the qubit, since decoherence effects in general do not~\cite{Wal03}. Let's therefore consider a superconducting ring that consists of two parts with different cross sections perpendicular to the flow, as shown in figure~\ref{tube}. By independently varying the cross sectional areas of the two regions we can then study the dependence of the collapse time on the geometry of the qubit. To simplify the calculation we will assume that the volumes of both parts are equal, and we will only consider a superposed flow of blocks of supercurrent which start at $t=0$ with their centers of mass exactly at the transition point between large and small cross sections.
\begin{figure}[hbt]
\center
\includegraphics[width=0.9\columnwidth]{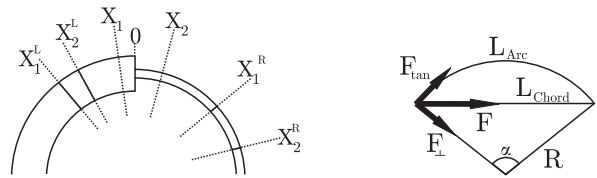}
\caption{Left: schematic representation of the separating blocks of current. The left and right side of block 1 are $X_1^L$ and $X_1^R$ respectively, and its centre of mass is at $X_1$. Right: diagram showing the orientation of forces in the flanks of the block of current.}
\label{defs}
\end{figure}

As in the case of the straight wire, the self energy will be due only to the non-overlapping flanks of the superposed pieces of supercurrent from $t=0$ until $t=t_1$. The distance between the centers of mass of the two copies, measured along the perimeter of the circularly shaped qubit, as depicted in figures~\ref{tube} and \ref{defs}, can straightforwardly be written as
\begin{eqnarray}
L_{\text{arc}} = \frac{1}{2} \left[ \left( \frac{A_1}{A_2} -1 \right) x_1 + \left( 1 - \frac{A_2}{A_1} \right) x_2 + \frac{A_1 + A_2}{A_1 A_2} \frac{M}{\rho} \right] \nonumber
\end{eqnarray}
where $M$ is the mass of the block that is to be superposed, and $A_1$ and $A_2$ are the cross sections perpendicular to the current flow in the different parts of the qubit.
The gravitational force between the two flanks does not act along the perimeter of the circle, but rather along the chord connecting the centers of the flanks. The part of the force along the tangent to the circle will have to be overcome in creating the superposition state, while the part of the force perpendicular to the circle will not contribute anything to the gravitational self energy and can be discarded (see figure~\ref{defs}).
From geometric considerations it is clear that if the currents moving in opposite directions are equal, then so are the masses of the flanks. Additionally, the centers of mass of the superposed blocks of current must then obey the relation $x_1=-\frac{A_2}{A_1} x_2$. Using this, it is easily shown that the gravitational self energy for $t<t_1$ is
\begin{eqnarray}
\Delta E &=& \xi G M_{\text{tot}}^2 \frac{1}{4 \pi N} \frac{ \cos \left( \frac{\pi}{N} \right)}{\sin^2 \left( \frac{\pi}{N} \right)} \frac{A_1 +A_2}{2 A_1 A_2} \frac{ \left( It / \rho_e \right)^2}{R^3}
\end{eqnarray}
where $\rho_e$ is the charge density of the supercurrent, $M_{\text{tot}}$ the total mass of the supercurrent around the full ring and $N \equiv M_{\text{tot}} / M$ is the number of patches of supercurrent needed to cover the whole ring.
In terms of the dimensionless quantities $X \equiv t \frac{I}{\rho_e R} \frac{A_1 +A_2}{2 A_1 A_2}$ and $Y \equiv \Delta E \frac{R}{\xi G M_{\text{tot}}^2} \frac{A_1 +A_2}{2 A_1 A_2}$, this equation reduces to
\begin{eqnarray}
Y = \frac{1}{4 \pi N} \frac{ \cos \left( \frac{\pi}{N} \right)}{\sin^2 \left( \frac{\pi}{N} \right)} X^2 &~& \text{for }~ 0 \leq X \leq \frac{\pi}{N}.
\end{eqnarray}

\begin{figure}[bt]
\center
\includegraphics[width=0.6\columnwidth]{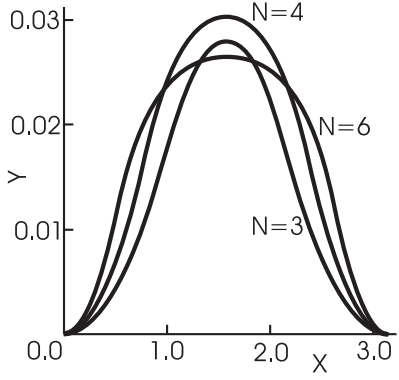}
\caption{The dimensionless energy $Y$ as a function of the dimensionless time $X$. The different curves represent different sizes of the block of supercurrent under consideration (parametrized by $N$).}
\label{XYcurrent}
\end{figure}

After the two copies of the block of current have completely detached from one another, the energy added to them during further separation is just the normal Newtonian self energy of two separated masses $\Delta E = \xi G M^2 \left( 1/x_0 - 1/x \right)$. Expressing this in terms of dimensionless units as well, and realizing that after the blocks have rounded a quarter of the ring they will start to come together again, we can write the full expression for the self energy during half a revolution as \index{self energy ! gravitational}
\begin{eqnarray}
\begin{array}{rll}
Y = & \frac{1}{4 \pi N} \frac{ \cos \left( \frac{\pi}{N} \right)}{\sin^2 \left( \frac{\pi}{N} \right)} X^2 & \text{for }~ 0 \leq X \leq \frac{\pi}{N} \nonumber \\
Y = & \frac{\pi}{4 N^3} \frac{ \cos \left( \frac{\pi}{N} \right)}{\sin^2 \left( \frac{\pi}{N} \right)}  & \nonumber \\
 & + \frac{1}{2 N^2} \left[ \frac{1}{\sin \left(\frac{\pi}{N}\right)} - \frac{1}{\sin \left( X \right)} \right] & \text{for }~ \frac{\pi}{N} \leq X \leq \frac{\pi}{2} \nonumber \\
Y\left( X\right)  = & Y\left( \pi - X \right) & \text{for }~ \frac{\pi}{2} \leq X \leq \pi. \nonumber
\end{array} \\
~
\label{QubitE}
\end{eqnarray}
This form of the gravitational self energy has been plotted in figure~\ref{XYcurrent} for different values of $N$.

If the current has not collapsed after half a revolution then the superposed blocks will be exactly on top of each other again, and thus stable with respect to gravity. We therefore only need to consider the current up to $X = \pi$, because if it has not collapsed by then, the supercurrent will be able to outrun the gravitational collapse process forever.

\begin{figure}[hbt]
\center
\includegraphics[width=0.9\columnwidth]{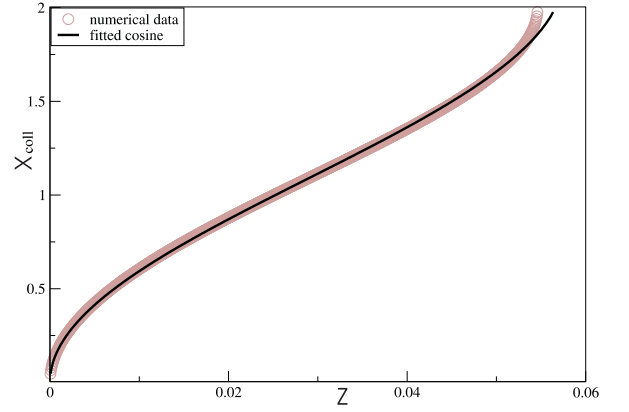}
\caption{The dimensionless reduction time $X_{\text{QSR}}$ as a function of the dimensionless current $Z$. The fit is the line defined by equation~\eqref{cosfit}.}
\label{fit}
\end{figure}

To convert the self energy in~\eqref{QubitE} to a time scale for gravity-induced quantum state reduction, we again need to look for the block of supercurrent which will first force the entire qubit to loose coherence. If we define $Y^{\text{max}}$ as the maximum $Y$ with respect to $N$ for a given $X$, then the condition for the reduction time to equal the inverse of the gravitational energy becomes, in dimensionless units
\begin{eqnarray}
X_{\text{QSR}} Y^{\text{max}} = Z,
\end{eqnarray}
with $Z \equiv I \frac{\hslash}{\xi G M_{\text{tot}}^2 \rho_e} \frac{A_1 +A_2}{2 A_1 A_2}$. We can then trace out the behavior of $X_{\text{QSR}}$ as a function of the 'current' $Z$. The curve thus found is depicted in figure~\ref{fit}. As it turns out, there is a very good and simple fit of the curve, given by
\begin{eqnarray}
Z = b \left[ 1 - \cos \left( \frac{\pi}{a} X_{\text{QSR}} \right) \right].
\label{cosfit}
\end{eqnarray}
with the fitting parameters $a=2.17$ and $b = 0.029$.

To come to the final expression for the collapse time in terms of the applied current and the geometrical parameters of the flux qubit, we reinsert dimensionful units, and invert the expression for $Z\left(X_{\text{QSR}}\right)$ to find
\begin{eqnarray}
t_{\text{QSR}} \hspace{-5pt} &=& \hspace{-5pt} \frac{a}{\pi} \frac{\rho_e}{I} \frac{2 A_1 A_2}{A_1 + A_2} R \cdot \nonumber \\
&& \hspace{-25pt} \arccos \left[ 1 - \frac{1}{b} \frac{\hslash}{\xi G m_e^2} \left(\frac{q_e}{2 \pi \rho_e}\right)^2 \frac{I}{\rho_e} \left( \frac{A_1 +A_2}{2 A_1 A_2} \right)^3 \frac{1}{R^2} \right]. \nonumber
\end{eqnarray}
Here $q_e$ and $m_e$ are the charge and mass of an electron. Notice that in the limit $R \to \infty$ the functional form of this expression reduces to that of equation~\eqref{mins}. (To be precise, the limit $R \to \infty$ yields $t_{\text{QSR}} = a /(\sqrt{2b} ~ \pi^2) ~ \sqrt{\hslash / G (\rho A)^2 v} = 0.91 \sqrt{\hslash / G (\rho A)^2 v}$. The difference in the prefactor from the earlier result is due to the fact that in the ring the collapse can also occur as the two copies of a piece of supercurrent are moving toward each other. If we consider only the self energy for pieces of supercurrent in one half of the ring, then the prefactor would indeed come out to be one.) From this expression it is immediately clear that if the current becomes large enough to make the argument of the arccosine smaller than $-1$, then there will be no quantum state reduction. This signifies the point at which the current is so fast that it can move a block of current of any size all the way around the (half) loop of the qubit before the gravitationally induced energy uncertainty has had a chance to initiate the quantum state reduction process. If $I$ is smaller than that, then the reduction is inevitable, and the corresponding timescale is given by $t_{\text{QSR}}$.

In order to be able to measure the time $t_{\text{QSR}}$, we will need to make it as small as possible. After all, the gravity-induced quantum state reduction must occur before any of the normal processes of decoherence has had a chance to destroy the superposition state. The difference between $A_1$ and $A_2$ does not seem very useful in that regard. However, the dependence of $t_{\text{QSR}}$ on $\left( A_1 + A_2 \right)/ 2 A_1 A_2$ does provide an additional test to see if we really are dealing with the effects of gravity, rather than just environmental decoherence.

Filling in the values for the natural constants, the condition for the current not to outrun its own reduction process becomes
\begin{eqnarray}
\frac{I}{R^2 A^3} &\leq& \xi 10^{27},
\end{eqnarray}
with $A=A_1=A_2$. The corresponding quantum state reduction time is given by
\begin{eqnarray}
t_{\text{QSR}} &\simeq& 10^8 \frac{A R}{I}.
\end{eqnarray}
The maximum contemporary coherence time of superconducting flux qubits like the ones used in the experiments in Delft~\cite{Wal00,Chiorescu00}, is about $100$~ns. The maximum current through the Josephson junction in the qubit is of the order of $1$~$\mu$A. To be able to trap just one fluxoid in the qubit, the radius is also limited to about $1$~$\mu$m, so that even if we could construct a qubit with the somewhat extravagant cross sectional area of $1$~x~$100$~$\mu$m$^2$, then still we would need to shield the superposition state from the environment for up to $10$ seconds; a factor $10^{8}$ above the presently feasible coherence time.

\subsection{Summary} \noindent
Is it possible to address empirically the two greatest mysteries of fundamental physics
(quantum gravity, wave function collapse) with the relatively modest means of modern
condensed matter- or cold atom experimentation? The analysis we presented here is
in this regard not particularly optimistic. Taking the Penrose hypothesis serious, the divide
between microscopic physics and the macroscopic world is still many orders of magnitude
away  from what is at present feasible in the laboratory. However, the history of physics is
littered with examples where the unimaginable has become daily routine because of the
role played by serendipity in the experimental pursuit. The real intention of this paper is
actually nothing else than an attempt to get Penrose's gravitational collapse on the bench
mark list shared by the various communities that are pursuing 'fat' quantum states in the
laboratory. The crucial aspect is the factor mass: in condensed matter and atomic physics
this is usually ignored because gravity is a very weak force. However, the essence of Penrose's vision is that it might exert remarkable powers and this should be in the forefront of everybody's
mind.

Let us view it as a contest. What are the strengths and weaknesses of the various approaches?
The 'nano-mechanical' approach has the virtue that it uses rigid bodies formed from
heavy atoms and it is trivial to imagine massive bodies being in spatial superposition -- the
Schr\"odinger cat. However, the price to pay is that such 'mechanical' states talk too easily to
the outside world with the effect that conventional decoherence obscures the view on the
gravitational collapse. The best way to isolate fat quantum states from the environment is
to invoke instead superfluid or superconducting rigidity. The Bose condensates formed from
cold atoms are in this regard unbeatable, and they have as an extra advantage that the
building material (atoms) is quite heavy. However, measured in units of the Penrose scale these
condensates are at present way too microscopic to be of any use, while it is far from clear
how to create true Schr\"odinger cat states.

A conceptual innovation introduced in this paper is our claim that not only mass-density-, but
also {\em mass current} superpositions should be sensitive to Penrose's gravitational
collapse. As we argued, this implies that the quantum bits constructed from superconducting
fluxoids which are now routinely produced, represent actually right now the closest approach
to the Penrose scale. In terms of the electrical currents involved (microamps) they are truly
macroscopic, but in terms of the Penrose dimensions these fall short because the mass of
the electron is tiny.

The bottomline is that in order to optimize the conditions one would like to invoke: (a) heavy
building blocks, i.e. atoms and not electrons, (b) it should be straightforward to orchestrate
the building blocks in large numbers and create Schr\"{o}dinger cat like states (the problem of the cold atoms),  (c) superfluid/superconducting
order is the way of choice in order to avoid decoherence, (d)  given superfluid  phase rigidity,
mass current superpositions are much more natural than mass density superpositions. In view of these conditions, superfluid Helium seems to be the ideal material for constructing this type of experimental test: it consists of atoms, it can easily be manipulated and it has the superfluid order and phase rigidity which in principle allows the superposition of superfluid flows. The only question is how to construct a true Schr\"{o}dinger cat state of superfluid Helium flows?

\subsection{Acknowledgements} \noindent
We acknowledge stimulating discussions with R. Penrose, P.H. Kes,
J. van den Brink, N.R. Cooper and J.C. Davis.


\end{document}